\documentclass[reprint, amsmath,amssymb, aps,]{revtex4-2} 	

\usepackage[dvips]{graphicx}
\usepackage{dcolumn}								
\usepackage{bm}
\usepackage{bm,amsmath}
\usepackage{braket}
\usepackage{ulem}
\usepackage{here}
\usepackage{float}

\usepackage{flushend}
\usepackage{balance}

\usepackage{silence}
\usepackage{color}

\DeclareRobustCommand{\YMdel}{\bgroup\markoverwith{\textcolor[rgb]{0.1, 0.5, 0.1}{\rule[.5ex]{2pt}{0.4pt}}}\ULon}

\DeclareRobustCommand{\HHdel}{\bgroup\markoverwith{\textcolor[rgb]{0.1, 0.1, 0.9}{\rule[.5ex]{2pt}{0.4pt}}}\ULon}

\begin{document}

\title{Fast Simulated Annealing inspired by Quantum Monte Carlo}
\author{Kiyotaka Murashima}
\email{murashima-kiyotaka@sei.co.jp}
\affiliation{
  R$\mathrm{\&}$ \hspace{-2.0mm}D Planning\hspace{0.5mm}$\mathrm{\&}$Administration Division, \nolinebreak Sumitomo Electric Industries, Ltd.
}%
\date{\today}
\begin{abstract}
Recently, much attention is being paid on Quantum Computing(QC) and some technologies are now using it in practice. 
An Ising spin model of interacting neighboring spins has been eagerly investigated along with the progress in QC.
Quantum Monte Carlo(QMC) is commonly used in such simulations but QMC, as a heuristic approach, 
has great difficultity in that it takes much time to search its minimum energy.
It mainly depends on the existence of a trotter layer derived from Suzuki-Trotter decomposition.
In this paper, I propose a new approach that changes the order 
among summing all the energy of spin configurations over the trotter layer and minimizing the energy.
Although the new method isn't rigorous, correct answers are obtained.
Its validity and advantage are also discussed in comparison with conventional Simulated Annealing(SA).
\end{abstract}
\pacs{Valid PACS appear here}
\maketitle

\section{Introduction}
Quantum annealing(QA) was first proposed by T. Kadowaki and H. Nishimori \cite{Kadowaki_1998}, 
and much theoretical investigation has been done using an Ising spin model.
Especially, Path-Integral Monte Carlo \cite{Martonak_2002} is a basic approach to calculate the behavoir of QA.
In calculating, the Hamiltonian is divided into two terms
that obeys a parallel and a perpendicular magnetic field to the spin axis.
As these terms don't commute each other, calculation will be done by Suzuki-Trotter decomposition with finite time intervals.
The spin configuration arisen from such a procedure is recognized as a trotter layer.
Much investigation has been done to culculate faster and more accurate \cite{Savard_2016} \cite{Muthumala_2020}.

SA is also a heuristic approach to solve optimization problems.
While SA occationally allows updating the state that has a higher energy,
SA is successful in not being trapped on local minimum.
The transition probability to the higher energy is tuned by changing annealing temperature.
But it hardly gets the correct answer of NP-hard problem.
So some improvements, for example, parallel tempering \cite{Wang} and so on, have been applied.

Under these circumstances, QA is thought to be able to calculate the optimization problems much faster.
But Simulated Quantum Annealing(SQA), based on the QMC, is not fast \cite{King}.
The reason is why it needs to sum all the energy of the trotter layer and to search all the spin configurations 
that minimize the total energy.
Furthermore, the fact also prevents SQA from quick convergence that the trotter number should be large.
Then, a new approach is proposed to overcome such difficulities, in section 2.
And in section 3, the fact is shown that it leads to correct answers, although it isn't rigorous mathematically.
Finally, its validity and advantagous points are argued in section 4, more precisely.

\section{Calculation algorithm}
First of all, let us briefly review QMC to solve 2D Ising spin model.
After that, I will detail a new approach.

\subsection{Quantum Monte Carlo}
Hamiltonian of the system has two terms, which are diagonal ($H_d$) and off-diagonal ($H_o$)
and they don't commute each other \cite{Martonak_2002}.
\begin{align}
    H  = - (H_d + H_o) = - \sum_{i<j} J_{ij} \hat{\sigma}_{z}^{i} \hat{\sigma}_{z}^{j} - \sum_{i} \mathit{\Gamma} \hat{\sigma}_{x}^{i}
\end{align}
where $\hat{\sigma}_{z}^{i}$, $\hat{\sigma}_{z}^{j}$ is Pauli z-matrix corresponding to the $i$-th and the $j$-th spin
and $\hat{\sigma}_{x}^{i}$ is Pauli x-matrix corresponding to the $i$-th spin.
The magnitude of the transverse magnetic field is denoted by $\mathit{\Gamma}$.

Next, the optimization problem is to find out the spin configuration that maximizes distribution function $Z$, below;
\begin{align}
    Z  = \mathrm{Tr} ( e^{-\beta H} )
        = \sum_{\psi_0} \langle \psi_0 | e^{-\beta H} |\psi_0 \rangle
\end{align}
where $|\psi_0 \rangle$ is wavefunction of spins 
and $\beta =1/ k_B T$ ($k_B$ is Boltzmann constant and $T$ is temperature).
Suzuki-Trotter decomposition is applied and $(2P-1)$ Identity Operators are inserted, then we get,
\begin{align}
    Z  &= \langle \psi_0 | e^{-\beta (\frac{H_d}{2P} + \frac{H_o}{2P})^{2P}} |\psi_0 \rangle \nonumber\\
        &= \sum_{\psi_0, \cdot \cdot \cdot ,\psi_{2P-1}}
              \langle \psi_0 | e^{-\beta \frac{H_d}{2P}} |\psi_{2P-1} \rangle
              \langle \psi_{2P-1} | e^{-\beta \frac{H_o}{2P}} |\psi_{2P-2} \rangle \nonumber\\
             &  \ \ \ \ \ \ \ \cdot \cdot \cdot \ \ 
              \langle \psi_2 | e^{-\beta \frac{H_d}{2P}} |\psi_1 \rangle
              \langle \psi_1 | e^{-\beta \frac{H_o}{2P}} |\psi_0 \rangle \nonumber\\
        &= \sum_{\psi_0, \cdot \cdot \cdot ,\psi_{P-1}}
              \langle \psi_0 | e^{-\beta \frac{H_d}{2P}} |\psi_0 \rangle
              \langle \psi_0 | e^{-\beta \frac{H_o}{2P}} |\psi_{p-1} \rangle \nonumber \\
             &  \ \ \ \ \ \ \ \cdot \cdot \cdot \ \  
              \langle \psi_1 | e^{-\beta \frac{H_d}{2P}} |\psi_1 \rangle
              \langle \psi_1 | e^{-\beta \frac{H_o}{2P}} |\psi_0 \rangle .
\end{align}
In the last expression, we use the fact that $e^{-\beta \frac{H_d}{2P}}$ is a diagonal operator.

The second term can be calculated furthermore, and finally the energy of this model can be expressed as classically:
\begin{align}
    E =& \sum_{k=0}^{P-1} \left (
                             \sum_{i<j} J_{ij} s_{z}^{i, k} s_{z}^{j, k} + \sum_{i} J_{\perp} s_{z}^{i, k} s_{z}^{i, k+1}
                             \right )  \\
    & \ \ \ \ J_{\perp} = - \frac{P T}{2} \mathrm{ln} \left (
                                                        \mathrm{tanh} \frac{\mathit{\Gamma}} {P T}
                                                        \right )   \nonumber
\end{align}
where $s_{z}^{i}$ denotes a classical spin, $\{\pm 1\}$.

\begin{figure}[t]
\includegraphics[keepaspectratio,scale=0.4, bb=-40 0 700 660]{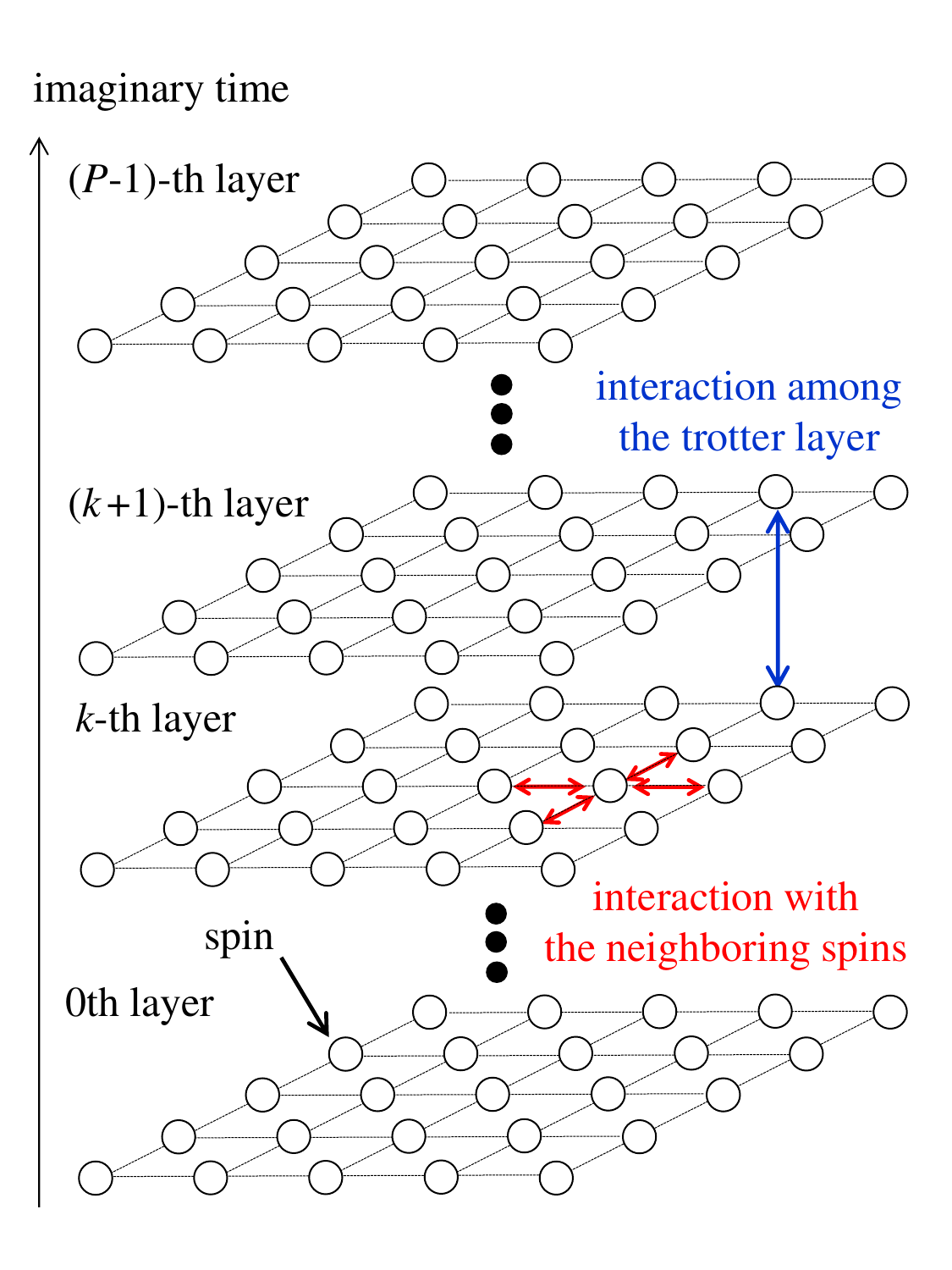}
\caption{
           Schematic picture of the trotter layer is shown.
           Trotter number $P$ should be large
           and the spin configuration expands over the imaginary time.
           In calculations, all the spin configurations have to be consistently decided
           by the interaction Hamiltonian.
}
\label{fig:trotter}
\end{figure}

As shown in Fig.\ref{fig:trotter}, 2D Hamiltonian turns out to be $3(=2+1)$ dimensional Hamiltonian 
having a new axis of imaginary time.
Additionally, the trotter layer has to meet a periodic boundary condition, such as $s_{z}^{i, 0} = s_{z}^{i, P}$.
Under these circumstances, the spin configuration that minimizes the energy in Equation (4) is searched.
Sometimes, there are some curious problems, 
for examples, "worldline" \cite{Ammon_1997} and "sign problem" \cite{Henelius_2000}.
But in this paper, such complicated problems are not discussed, as we only focus on the minimum energy.

\subsection{A New Algorithm inspired by QMC}
In previous, a classical expression of 3D Hamiltonian was obtained.
But, let us go back to express it in quantum ways.
Using the wavefunction of the $k$-th trotter layer, $|\psi_k \rangle$,
it can be understood as in Equation (5).
\begin{align}
    E = \sum_{k=0}^{P-1}  \ \biggl (&
                              \langle \psi_k |
                              \sum_{i<j} J_{ij} \hat{\sigma}_{z}^{i, k} \hat{\sigma}_{z}^{j, k}
                              | \psi_k \rangle \nonumber \\
       & \ \ \ +
                              \langle \psi_k |
                              \sum_{i} J_{\perp} \hat{\sigma}_{z}^{i, k} \hat{\sigma}_{z}^{i, k+1}
                              | \psi_{k+1} \rangle
                              \biggl )
\end{align}

\begin{figure}[t!]
\includegraphics[keepaspectratio,scale=0.40, bb=-35 0 680 680]{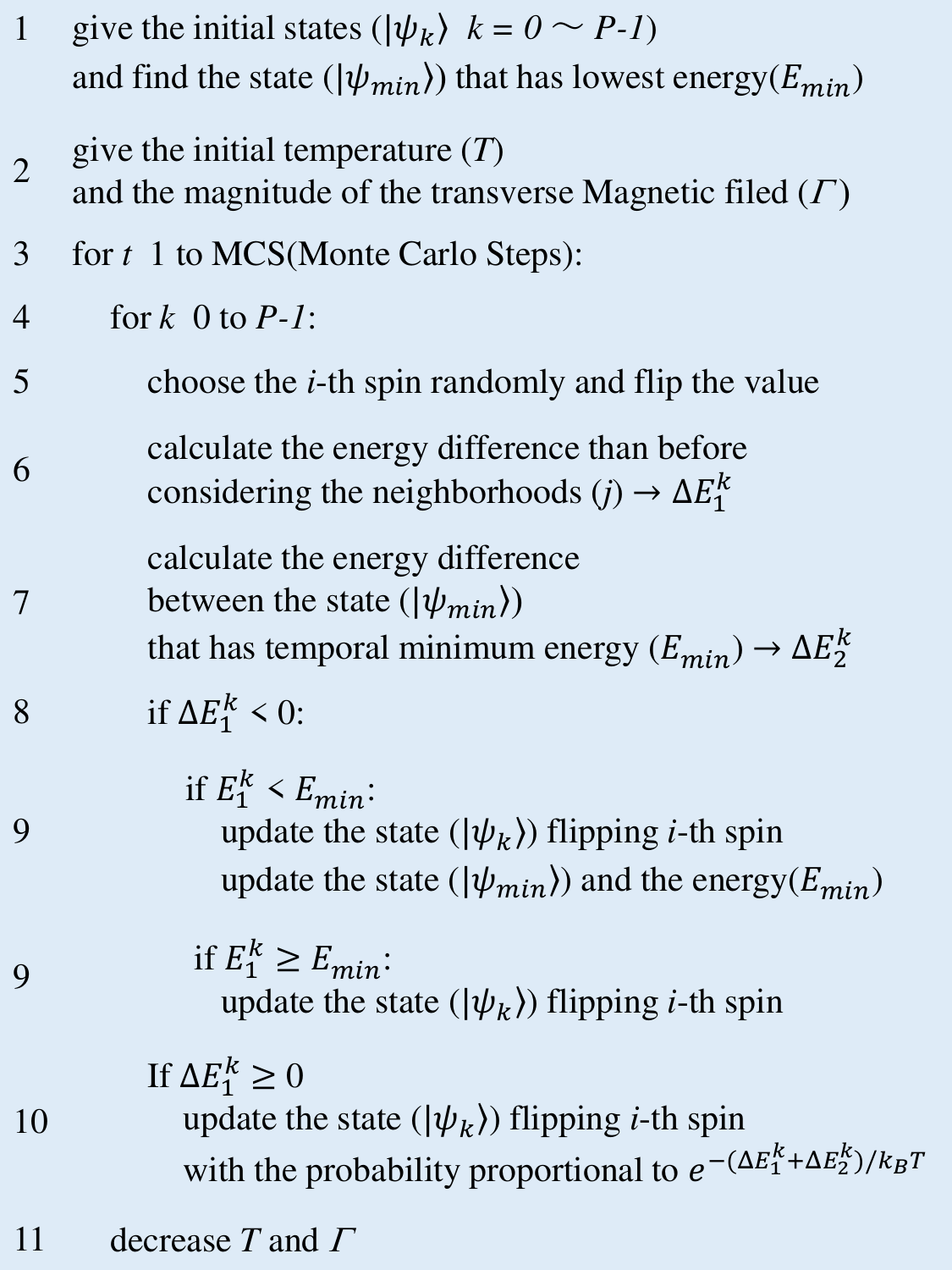}
\caption{
            Flowchart of the new algorithm is shown. The Conventional QMC has to sum all the energy of the trotter layer,
            after randomly updating the spins. 
            It levels the energy, so that the procedure prevents quick convergence.
            The new approach skips it and each trotter layer imitates the spin configuration of a temporal minimum energy.
}
\label{fig:flowchart}
\end{figure}

So, searching the spin configuration that has minimum energy is explicitly expressed as below:
\begin{align}
    & \hspace{-20mm}  \mathrm{argmin} \ E \nonumber \\
    & \hspace{-20mm} =\underset{\psi_{0}, \cdot \cdot \cdot ,\psi_{p-1}}{\mathrm{argmin}}
                              \biggl (
                              \langle \psi_{P-1} |
                              \sum_{i<j} J_{ij} \hat{\sigma}_{z}^{i, P-1} \hat{\sigma}_{z}^{j, P-1}
                              | \psi_{P-1} \rangle \nonumber \\
       &  \ \ \ \ \ \ \ \ \ \ \ +
                              \langle \psi_{P-1} |
                              \sum_{i} J_{\perp} \hat{\sigma}_{z}^{i, P-1} \hat{\sigma}_{z}^{i, 0}
                              | \psi_{0} \rangle \nonumber \\
       \ \ \ \ \ \ \ \ \ \ \ \ \ \ \ \ \ \ \cdot \nonumber \\
       \ \ \ \ \ \ \ \ \ \ \ \ \ \ \ \ \ \ \cdot \nonumber \\
       \vspace{0.9cm}
      & \hspace{-20mm} \ \ + 
                              \langle \psi_{1} |
                              \sum_{i<j} J_{ij} \hat{\sigma}_{z}^{i, 1} \hat{\sigma}_{z}^{j, 1}
                              | \psi_{1} \rangle
      +
                              \langle \psi_{1} |
                              \sum_{i} J_{\perp} \hat{\sigma}_{z}^{i, 1} \hat{\sigma}_{z}^{i, 2}
                              | \psi_{2} \rangle \nonumber \\
      &  \hspace{-20mm} \ \ + 
                              \langle \psi_{0} |
                              \sum_{i<j} J_{ij} \hat{\sigma}_{z}^{i, 0} \hat{\sigma}_{z}^{j, 0}
                              | \psi_{0} \rangle
      +
                              \langle \psi_{0} |
                              \sum_{i} J_{\perp} \hat{\sigma}_{z}^{i, 0} \hat{\sigma}_{z}^{i, 1}
                              | \psi_{1} \rangle
                              \biggl )
\end{align}


\clearpage

\begin{table*}[t!]
  \vspace{-0.4cm}
  \caption{Parameters and Calculation Results \lbrack CASE-1\rbrack}
  \label{table:data_type}
  \centering
  \begin{tabular}{|c||c|c|c||c|c|c||c|c|}
    \hline
    \ graph \ &\  node \ &\  edge  \ &\  BKC                  \ &\ trotter \ &\  initial          \ &\  trial times \ &\  ave. \ &\   the times of \  \\
            &             &          &\  (Best Known Cut) \  &\ number \           &\  temperature \ &               &       & reaching BKC \\
    \hline

    G9   & 800 & 19176 & 2054 & 150 & 0.0096 & 100 & 2052.0 & 55 \\
    G13  & 800 & 1600 & 582 & 150 & 0.0040 & 50 & 579.5 & 10 \\
    G18  & 800 & 4694 & 992 & 150 & 0.0050 & 50 & 989.0 & 2 \\
    G19  & 800 & 4661 & 906 & 150 & 0.0065 & 50 & 904.7 & 17 \\
    G20  & 800 & 4672 & 941 & 150 & 0.0065 & 50 & 941.0 & 50 \\
    G21  & 800 & 4667 & 931 & 150 & 0.0065 & 50 & 928.3 & 4 \\
    G34  & 2000 & 4000 & 1384 & 200 & 0.0050 & 50 & 1379.4 & 2 \\
    \hline
  \end{tabular}
  \label{tb:param}
\end{table*}

\begin{figure*}[t!]
 \vspace{-0.8cm}
 \centering
  \includegraphics[scale=0.7, bb=-5 0 800 250]{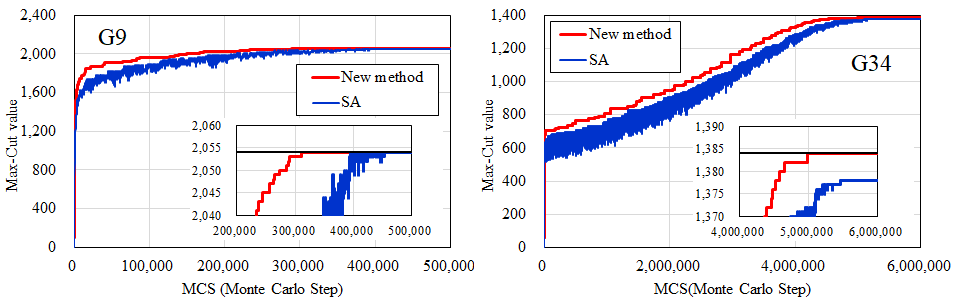}
  \caption{
             Convergence behaviors of G9 and G34 are shown.
             Both calculations by the new method have reached their BKC(Best Known Cut), although SA didn't for G34.
             Inset apparently shows that the new method could converge faster than SA.
}
\label{fig:graphs}
\end{figure*}

\begin{table*}[t!]
  \caption{Parameters and Calculation Results \lbrack CASE-2\rbrack}
  \label{table:data_type}
  \centering
  \begin{tabular}{|c||c|c|c||c|c|c||c|c|}
    \hline
    \ graph \ &\  node \ &\  edge  \ &\  BKC                  \ &\ trotter \ &\  initial          \ &\  trial times \ &\  ave. \ &\   the times of \  \\
            &             &          &\  (Best Known Cut) \  &\ number \           &\  temperature \ &               &       & reaching BKC \\
    \hline

    G9   & 800 & 19176 & 2054 & 150 & 0.0040 & 50 & 2047.0 & 6 \\
    G20  & 800 & 4672 & 941 & 150 & 0.0020 & 50 & 938.9 & 17 \\
    G34  & 2000 & 4000 & 1384 & 200 & 0.0010 & 50 & 1381.2 & 12 \\
    \hline
  \end{tabular}
  \label{tb:param2}
\end{table*}

\begin{figure*}[t!]
 \vspace{-0.8cm}
 \hspace{0.6cm}
 \centering
  \includegraphics[scale=0.7, bb=-5 0 800 250]{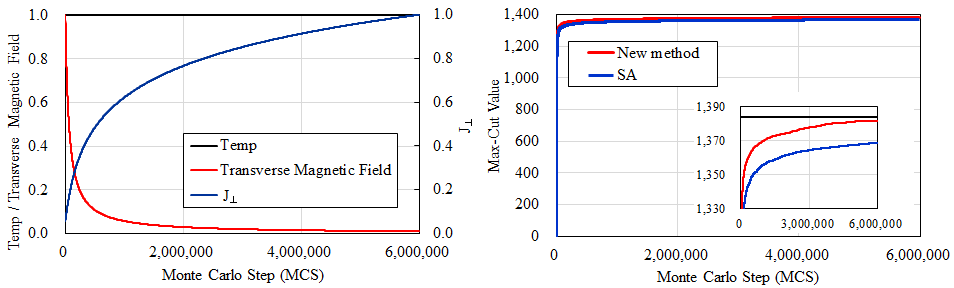}
  \caption{
             In calculation, temperature was constant and the transverse magnetic filed was controlled
             in order for $J_{\perp}$ to behave like in the left figure.
             The new method converged faster and reached better Max-Cut values than SA on average.
}
\label{fig:graphs2}
\end{figure*}

Now, let us change the order among summing all the energy over the trotter layer 
and finding out the spin configuration that minimize the energy.
Of course, it is not correct mathematically,
but if we only focus on the minimum energy, $\mathrm{argmin}$ could be applied step by step.
Based on the above idea, firstly $\mathrm{argmin}$ shall be applied to one side(ket vector) 
of the spin configuration of the second term.
After that, the energy-minimum state shall be searched.
Then, we finally get Equation (7).

\begin{align}
    & \hspace{-1.5mm}  \mathrm{argmin}  \ E \nonumber \\
    & \hspace{-1.5mm} = \underset{\psi_{k}}{\mathrm{argmin}} \ 
                              P \times \biggl (
                              \langle \psi_{k} |
                              \sum_{i<j} J_{ij} \hat{\sigma}_{z}^{i, k} \hat{\sigma}_{z}^{j, k}
                              | \psi_{k} \rangle \nonumber \\
       &  \ \ \ \ \ \ \ \ \ \ \ \ \  \ \ \ \ +
                              \langle \psi_{k} |
                              \sum_{i} J_{\perp} \hat{\sigma}_{z}^{i, k} \hat{\sigma}_{z}^{i, min}
                              \Bigl{\{} \underset{\psi_{k+1}}{\mathrm{argmin}} \ | \psi_{k+1} \rangle \Bigl{\}}
                              \biggl )   \nonumber \\
    & \hspace{-1.5mm} = \underset{\psi_{k}}{\mathrm{argmin}} \ 
                              P \times \biggl (
                              \langle \psi_{k} |
                              \sum_{i<j} J_{ij} \hat{\sigma}_{z}^{i, k} \hat{\sigma}_{z}^{j, k}
                              | \psi_{k} \rangle \nonumber \\
       &  \ \ \ \ \ \ \ \ \ \ \ \ \  \ \ \ \ +
                              \langle \psi_{k} |
                              \sum_{i} J_{\perp} \hat{\sigma}_{z}^{i, k} \hat{\sigma}_{z}^{i, min}
                              | \psi_{min} \rangle
                              \biggl ) 
\end{align}

In Equation (7), the problem comes to only search the spin configuration,
while imitating the temporal energy-minimum state among $P$ canditates.
In order to express it more explicitly, we use $\hat{\sigma}_{z}^{i, min}$ 
as a Pauli z-matrix corresponding to such an energy-minimum state, $| \psi_{min} \rangle$.
Flowchart of the new algorithm is shown in Fig.\ref{fig:flowchart}.

\section{Calculation Results}

\subsection{Comparison with SA}
We examined its validity by solving MAX-CUT problems, GSET \cite{GSET}.
It has been commonly used to estimate the performance of various algorithms \cite{Muthumala_2020} \cite{Benlic}, 
for its Best Known Cut(BKC) was well-investigated.

We calculated typical GSET with Python 3.10.7 and an Intel$^{\circledR}$ Core$^{\mathrm{TM}}$ i5-10210U Processor.
As for spin update, spin flip was done one by one at each trotter layer 
and the upper limit of Monte Carlo Steps(MCS) was set at 500,000 for 800 nodes and 6,000,000 for 2,000 nodes.
The other conditions are shown in Table-I.
Temperatures in Table-I are initial values and the initial magnitude of transverse magnetic field was set at 1.0.
At final MCS, both the temperature and the magnitude of the transverse magnetic field were decreased
to the initial values multiplied by $10^{-6}$.
And $J_{\perp}$ in equation (4) was a little bit adjusted, 
in order to increase the quantum effect and to be the same spin configuration in the end.

In Fig.\ref{fig:graphs}, convergence behaviors of the new algorithm and the conventional SA are shown.
The new algorithm reached BKCs for all cases.
For G9, 800 nodes, SA also reached BKC but the speed was slower.
And in the case of SA, the number of times reaching BKC was only 3. It was very few.
Furthermore, for G34, 2,000 nodes, the conventional SA couldn't reach BKC after 100 trials.
Of course, the comparison is quite unfair, because the new method could try $P$ times at once.
But it will not matter. The reason will be explained, later.

\subsection{Performance as QMC}
In the previous calculation, the temperature was changed 
in order to clearly show the difference between QMC and SA.
But such a condition doesn't reflect the true QA, because the annealing temperature is not enough high to be controled. 
Next, another calculation was done with a fixed temperature
and only the magnitude of the transverse magnetic field was tuned.
$\mathit{\Gamma}$ was changed inversely proportional to MCS.
Calculation results are shown in Table-I\hspace{-1.2pt}I 
and the convergence behaviors for G34 is shown in Fig.\ref{fig:graphs2}.

The new method and the SA curves were very similar, so the data were averaged over 50 trials.
Fig.\ref{fig:graphs2} clearly shows that the new algorithm converged faster than SA
and the possibility for QA to reach the solution is much higher than that of SA.

\section{Discussions}

\subsection{Searching Policy and its Advantages}
\begin{figure}[H]
\includegraphics[keepaspectratio,scale=0.4, bb=-30 0 150 670]{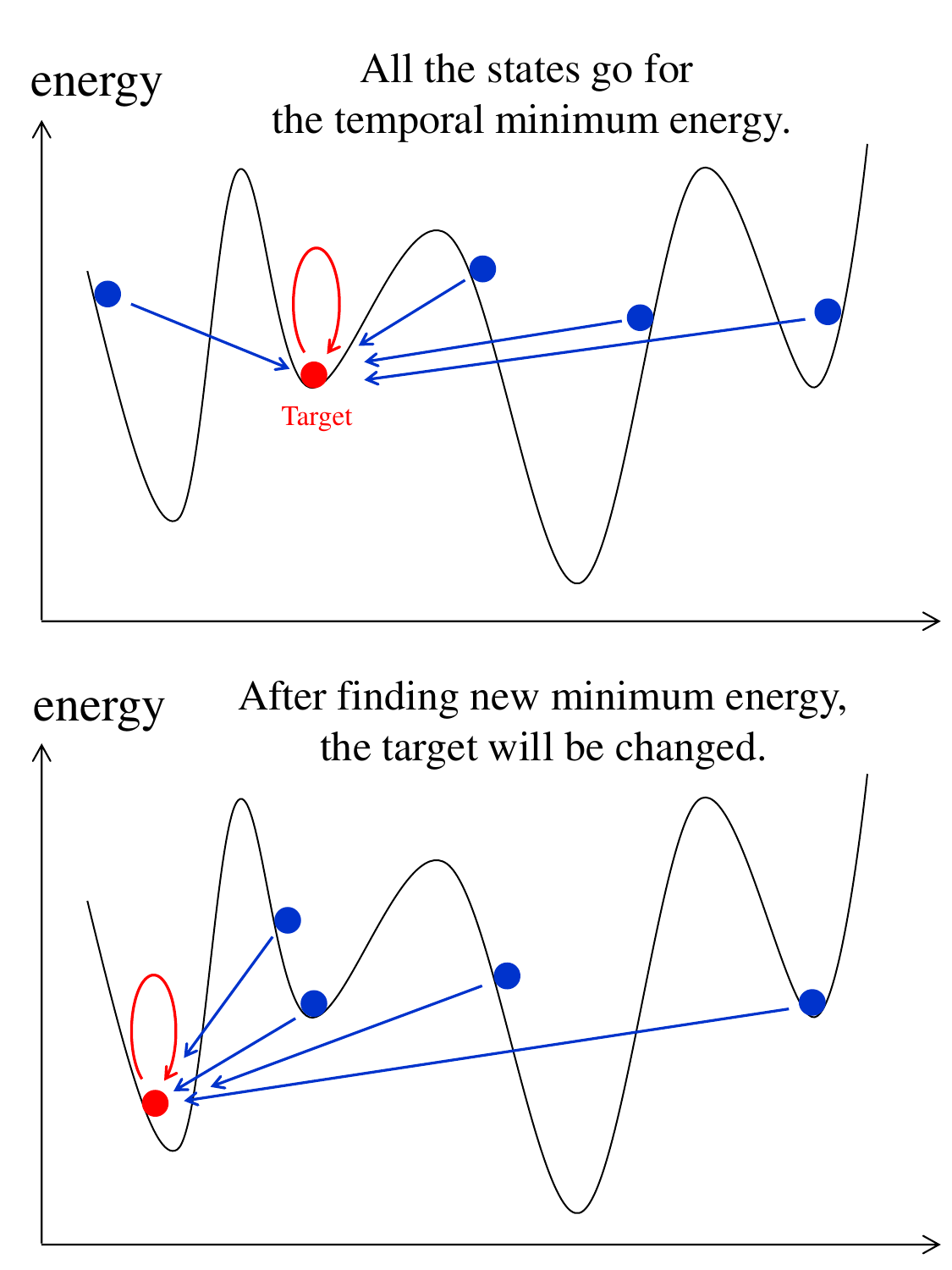}

\caption{
           Schematic picture is shown to search minimum energy with the new approach.
           All the trotter layers update the spin independenlty, but while imitating that of a temporal minimum energy.
           And once a new energy-minimum is found, the target state will be changed quickly.
           Like this way, every state is searching the minimum energy.
           The effect is tuned by the second term of Equation (7).
}
\label{fig:concept}
\end{figure}

Generally, the difference between SA and QMC is thought to be the existence of "Tunnelling effect".
In QMC, the state can jump a higher energy and quickly reach another lower energy.
The new method is the very picture that all the trotter layers co-exixt and are searching the minimum energy.
It can be also said that they are constantly competing, while searching their minimum energy.
The situation is shown in Fig.\ref{fig:concept}.

And interestingly, Equation (7) only needs the spin configuration 
that gives a temporal minimum energy $| \psi_{min} \rangle$.
Therefore, we can search such a state among $P$ candidates $| \psi_{k} \rangle$ independently.
Once we find out a new minimum energy, we only have to change the target and continue the calculation.
Such a so-called, parallel procedure, could be applied.
Nowadays, with Graphical Processing Unit(GPU) or Field Programmable Gate Array(FPGA),
such calculations would be easily performed and it might be much faster.

\subsection{Validity of New Algorithm}
As is thought in QMC, every spin configuration of the trotter layer shalll be same in the end.
In the new approach, every spin configuration goes for that of a temporal minimum energy
and such a state shall be constantly updated.
Therefore, the operating principle is kept the same.

\begin{figure}[H]
\includegraphics[keepaspectratio,scale=0.67, bb=-2 0 460 430]{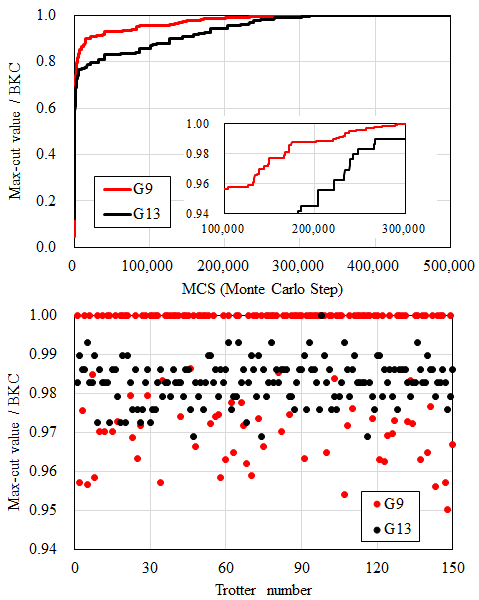}
\caption{
           The convergence behaviors of all the trotter layers are shown chaging the annealing temperature, CASE-1.
           As expected, finally 93 spin configurations out of 150 have the same energy for G9.
           This result proves that the concept shown in Fig. \ref{fig:concept} functions effectively.
}
\label{fig:trotter-G9}
\end{figure}

This is proven in Fig.\ref{fig:trotter-G9}, by the fact that 
almost all of the trotter layers finally had the same energy in the case of G9.
But in the case of G13, it seemed not to work efficiently.
It might be the reason why G13 has 3 different energy-lowest states, the solution was degenerated.
Such a behavior was also appeared in Fig.\ref{fig:trotter-G9}.
Instead of a steady increase of G9, the increase of G13 was quite gradual.
It reflected that the target state is often changed before convergence.

\section{Conclusions}

A new SA algorithm inspired by QMC was proposed.
Its validity was proven by the fact that it could reach BKC 
and the spin configuration of each trotter layer tended to be the same.
It couldn't be proven that the new algorithm would be faster than SA,
becauae caluculations were done without GPU nor FPGA,
But the posibility of it was shown theoretically.
I hope the new algorithm will help the other spin models easy to be handled.


\newpage

\end{document}